\documentclass[a4paper,iop,numberedappendix]{emulateapj}
\usepackage[colorlinks=true,linkcolor=blue,anchorcolor=blue,citecolor=blue,urlcolor=blue]{hyperref}
\usepackage{color}
\hypersetup{pdfauthor={Keiichi Umetsu}, pdftitle={Multi-Probe Cluster Lens Reconstruction}}
\usepackage{natbib}

\newcommand{\simgt}{\lower.5ex\hbox{$\; \buildrel > \over \sim \;$}}
\newcommand{\simlt}{\lower.5ex\hbox{$\; \buildrel < \over \sim \;$}}

\def\bc{\mbox{\boldmath $c$}}
\def\btheta{\mbox{\boldmath $\theta$}}

\def\bd{\mbox{\boldmath $d$}}
\def\bs{\mbox{\boldmath $s$}}

\lefthead{Umetsu}

\righthead{}

\begin{document}

\title{Model-Free Multi-Probe Lensing Reconstruction of Cluster Mass Profiles\altaffilmark{*}}
\author{Keiichi Umetsu\altaffilmark{1}} 
\altaffiltext{*}
 {Based in part on data collected at the Subaru telescope,
  which is operated by the National Astronomical Society of Japan.}
\altaffiltext{1}
 {Institute of Astronomy and Astrophysics, Academia Sinica,
  P.~O. Box 23-141, Taipei 10617, Taiwan.}

\begin{abstract}
Lens magnification by galaxy clusters induces characteristic spatial
 variations in the number counts of background sources, amplifying their
 observed fluxes and expanding the area of sky, the net effect of which,
 known as magnification bias, depends on the intrinsic faint-end slope
 of the source luminosity function.  The bias is strongly negative for
 red galaxies, dominated by the geometric area distortion, whereas it is
 mildly positive for blue galaxies, enhancing the blue counts toward the
 cluster center.   
We generalize the Bayesian approach of Umetsu et al. 
for reconstructing projected cluster mass profiles, by
 incorporating multiple populations of background sources for
 magnification bias measurements and combining them with complementary
 lens distortion measurements, 
effectively breaking the mass-sheet degeneracy and improving the
 statistical precision of cluster mass measurements.
The approach can be further extended  to include strong-lensing
 projected mass estimates, thus allowing for
 non-parametric absolute mass determinations in both the weak and strong
 regimes.  We apply this method to our recent CLASH lensing measurements 
 of MACS~J1206.2$-$0847, and demonstrate how combining multi-probe lensing
 constraints can improve the reconstruction of cluster mass profiles. 
 This method will also be useful for a stacked lensing analysis,
 combining all lensing-related effects in the cluster regime, 
 for a definitive determination of the averaged mass profile.
\end{abstract}
 

\section{Introduction} 
\label{sec:intro}

Galaxy clusters, the largest self-gravitating systems in the universe, 
represent a fundamental class of astrophysical objects,
which contain a wealth of information about 
the initial conditions of primordial density fluctuations,
the emergence and growth of nonlinear structure over cosmic time.
Clusters can therefore provide fundamental constraints on
the nature of dark matter (hereafter DM) \citep{Clowe+2006_Bullet},
alternative gravity theories \citep{Narikawa+Yamamoto2012},
and models of structure formation \citep{Allen+2008_fgas,Vikhlinin+2009_CCC3}.
   
The standard $\Lambda$ cold dark matter ($\Lambda$CDM) model
provides testable predictions for the structure and environment of
collisionless CDM  halos. 
In this context, $N$-body simulations have established an approximately
self-similar form for the spherically-averaged density profile
$\langle\rho(r)\rangle$ of quasi-equilibrium CDM halos 
\citep[][hereafter Navarro-Frenk-White, NFW]{1997ApJ...490..493N}
over a wide range of halo masses and radii, with some intrinsic variance
associated with assembly bias and dynamical structure of individual halos
\citep{Jing+Suto2000,Tasitsiomi+2004,Graham+2006,Faltenbacher+2009,Lapi+Cavaliere2009a,Navarro+2010,Gao+2012_Phoenix}.

Massive clusters act as powerful gravitational lenses, producing various
detectable effects via 
shifting, magnifying, and distorting the images of distant background
sources \citep{2001PhR...340..291B}.
Gravitational lensing thus offers a unique opportunity to 
study the underlying matter distribution in and around cluster-sized halos
\citep{Umetsu2010_Fermi,Kneib+Natarajan2011}, 
irrespective of the physical nature, 
composition, and state of lensing matter \citep{Okabe&Umetsu08},
providing a direct probe for testing well-defined predictions
\citep{Oguri+Hamana2011,Silva+2013}. 
Careful lensing work has shown that
the total mass profiles of clusters
exhibit a steepening radial trend with a clear profile curvature,
in overall agreement with the predicted form for the family of
CDM-dominated halos
\citep{2003A&A...403...11G,BTU+05,2007ApJ...668..643L,UB2008,2008MNRAS.386.1092L,Okabe+2010_WL,Umetsu+2010_CL0024,Sereno+Umetsu2011,Umetsu+2011,Umetsu+2011_stack,Umetsu+2012,Coe+2012_A2261,Newman+2012a,Newman+2012b}. 

The main attraction of cluster gravitational lensing is its ability to
make a model-independent determination of halo mass profiles\footnote{We
remind the reader that model dependence is unavoidable to some extent in
scientific analysis. In this work we define the term ``model
independent'' to refer to those methods without prior assumptions about the functional
form of the cluster radial profiles and mass distributions.}
over a wide range of cluster radii,
when the complementary effects of weak and strong lensing
are combined together
\citep{BTU+05,UB2008,Merten+2009,Zitrin+2010_A1703,Zitrin+2011_A383,Umetsu+2010_CL0024,Umetsu+2011,Umetsu+2011_stack,Umetsu+2012,Oguri+2012_SGAS}.
It has been demonstrated by our earlier weak-lensing work
\citep{BTU+05,UB2008,Medezinski+2010} that,
without adequate color information, the measured distortion
signal can be substantially diluted due to the contamination by unlensed cluster
members, leading to biased mass-profile measurements with
underestimated profile concentrations, 
underpredicting the observed Einstein radius from strong lensing.

Internal consistency of lensing measurements can be further tested by
measuring the complementary magnification effects.
Gravitational magnification can influence the observed surface
number density of background sources, enhancing their apparent fluxes
and expanding the area of sky  
\citep{1995ApJ...438...49B,UB2008,Hildebrandt+vanWaerbeke+Erben2009,Hildebrandt+2011,vanWaerbeke+2010,Rozo+Schmidt2010,Umetsu+2011,Umetsu+2012,Huff+Graves2011,Ford+2012,Schmidt+2012,Morrison+2012}.   
The former effect increases the observable number of sources above the
limiting flux, 
whereas the latter reduces the effective observing  
area in the source plane, thus decreasing the number of sources per solid angle.
The net effect is known as magnification bias, and depends on the
intrinsic faint-end slope of the source luminosity function.
 
In this paper we pursue the utility of all possible lensing information
available in the cluster regime, 
for making high-accuracy, model-free measurements of the cluster mass
profile from a joint likelihood analysis of  
multi-probe lensing observations.
This extends the Bayesian approach of \citet{Umetsu+2011}
based on the unique combination of 
weak-lensing distortion and {\it negative} magnification-bias
measurements, effectively breaking
degeneracies inherent in a standard weak-lensing analysis based on
shape information alone
\citep[see][]{1995A&A...294..411S,Schneider+2000}. 
The Bayesian method of \citet{Umetsu+2011} 
has been extensively used to reconstruct the projected mass
profile in a dozen clusters: A1689, A1703, A370,
Cl0024$+$17, RXJ1347$-$11 \citep{Umetsu+2011}, 
A383 \citep{Zitrin+2011_A383},
A2261 \citep{Coe+2012_A2261}, 
MACS~J1206.2$-$0847 \citep{Umetsu+2012}, 
MACS~J0416.1$-$2403 \citep{Zitrin+2013_M0416},
and
MACS~J0717.5$+$3745 \citep{Medezinski+2013}.
In all cases, we find a good
agreement between independent weak- and strong-lensing based 
mass profiles in the region of overlap.

Here, we shall make a full use of magnification-bias effects, by
extending source-count measurements into multiple populations of
background sources, probing a wider range of levels of magnification
bias.
The strong- and weak-lensing regimes contribute
quite similar logarithmic coverage of the radial profile
for a massive cluster acting as a super-critical lens
\citep{Umetsu+2011,Umetsu+2011_stack,Umetsu+2012}.  In this work,
we generalize the \citet{Umetsu+2011} method to combine the
central strong-lensing information with weak-lensing measurements
in a joint mass-profile analysis.



The paper is organized as follows. 
In Section~\ref{sec:basis} we briefly summarize the basic theory of cluster
weak gravitational lensing.  
In Section~\ref{sec:method} we outline our comprehensive methods for
obtaining projected cluster mass profiles from multi-probe cluster
lensing observations.
%
In Section~\ref{sec:app} we apply our methodology to our recent CLASH
lensing observations of the massive cluster MACS~J1206.2$-$0847 (hereafter
MACS1206) at $z=0.439$
\citep{Postman+2012_CLASH,Zitrin+2012_M1206,Umetsu+2012}
to derive a joint mass profile solution from the combined
strong-lensing,
weak-lensing distortion, positive and negative magnification-bias
measurements, 
demonstrating how complementary multi-probe cluster lensing measurements
can improve the reconstruction of cluster mass profiles.
Finally, summary and discussions are given in Section~\ref{sec:summary}. 
 
Throughout this paper, we use the AB magnitude system, and
adopt a concordance $\Lambda$CDM cosmology
with $\Omega_{m}=0.3$,  
$\Omega_{\Lambda}=0.7$, and $h\equiv H_0/(100\, {\rm km\, s^{-1}\,
Mpc^{-1}})=0.7$. Errors represent a confidence level of 68.3\% ($1\sigma$) unless otherwise
stated.

\section{Cluster Weak Lensing Theory}
\label{sec:basis}

\subsection{Convergence, Shear, and Magnification}

The central quantity in this work is the lensing convergence
$\kappa$.
For cluster lensing, it is expressed as 
$\kappa(\btheta)=\Sigma(\btheta)/\Sigma_{\rm crit}$,
namely the projected mass density $\Sigma(\btheta)$
in units of the critical surface mass density,
\begin{equation} 
\label{eq:sigmacrit}
\Sigma_{\rm crit} = \frac{c^2}{4\pi G D_l} \beta^{-1};
\ \ \ \beta(z) \equiv D_{ls}/D_s,
\end{equation}
where $D_l$, $D_s$, and $D_{ls}$ are the proper angular diameter
distances from the observer to the lens ($l$), 
the observer to the source ($s$), and the lens to the source, respectively.
The distance ratio $\beta(z)$ represents the geometric strength of cluster lensing
for a source at redshift $z$; $\beta(z)=0$ for unlensed objects, $z\le z_l$.

Weak gravitational lensing is responsible for the magnification and
shearing of the images of background sources
\citep{2001A&A...378..361B,Umetsu2010_Fermi,Kneib+Natarajan2011}.
The image deformation is described by 
the $2\times 2$ Jacobian matrix ${\cal A}$ of the lens mapping,
\begin{equation}
{\cal A} = (1-\kappa){\cal I} -\gamma_1\sigma_3-\gamma_2\sigma_1,
\end{equation}
with ${\cal I}$ the identity matrix,
$\sigma_a$ ($a=1, 2, 3$) the Pauli matrices,  
and $(\gamma_1,\gamma_2)$ the components of the complex gravitational shear $\gamma=\gamma_1+i\gamma_2\equiv |\gamma|e^{2i\phi}$ 
with spin-2 rotational symmetry \citep[e.g.,][]{2001PhR...340..291B}.  
%

The lens magnification is given by the inverse Jacobian determinant,
$\mu(\btheta)\equiv 1/|{\rm det}{\cal A}(\btheta)|=1/|(1-\kappa)^2-|\gamma|^2|$. 
%
%
In the  weak-lensing limit $\kappa,|\gamma|\ll 1$,
the shear alone induces a quadrupole anisotropy of the  
background images, which can be observed from ellipticities 
of background galaxy images~\citep{1995ApJ...449..460K}.
In general, the observable quantity for quadrupole weak lensing
is not $\gamma$ but the {\it reduced} gravitational shear
$g(\btheta)=\gamma(\btheta)/[1-\kappa(\btheta)]$
in the subcritical-lensing regime where ${\rm det}{\cal A}>0$
(or $1/g^*$ in the negative parity region with ${\rm det}{\cal A}<0$ and $|g|>1$).

\subsection{Mass-sheet Degeneracy}
\label{subsec:sheet}

The observable distortion field $g(\btheta)$ is invariant under the following
global linear transformation:
\begin{equation}
\label{eq:invtrans}
\kappa(\btheta) \to \lambda \kappa(\btheta) + 1-\lambda, \ \ \ 
\gamma(\btheta) \to \lambda \gamma(\btheta)
\end{equation}
with an arbitrary scalar constant $\lambda\ne 0$,
which is known as the mass-sheet or steepness degeneracy 
\citep[see][]{1995A&A...294..411S,Saha+Williams2006,Liesenborgs2012}.
This global transformation is equivalent to scaling 
${\cal A}(\btheta)$ with $\lambda$, 
$\cal {A}(\btheta) \to \lambda {\cal
A}(\btheta)$, hence leaving the critical curves ${\rm det}{\cal
A}(\btheta)=0$ in the strong-lensing regime invariant.  
%
This degeneracy can be broken,
for example,\footnote{Alternatively, the constant $\lambda$ can be determined such that
the $\kappa$ averaged over the outermost cluster region
vanishes, if a sufficiently wide sky coverage is available.
Or, one may constrain $\lambda$ such that the enclosed mass
within a certain aperture is consistent with cluster mass
estimates from some other observations
\citep[e.g.,][]{Umetsu+Futamase1999}. 
}
by measuring the magnification effects (Section~\ref{subsec:magbias}),
because $\mu$ transforms as $\mu(\btheta)\to \lambda^2\mu(\btheta)$.

\subsection{Source Populations and Redshift Distributions}
\label{subsec:pop}

For statistical weak-lensing measurements, 
we consider populations ($\alpha=1,2,...$) of source galaxies with respective redshift distribution
functions $N_{(\alpha)}(z)$.\footnote{In general, 
we apply different size, magnitude, and color cuts in source selection for measuring the lens distortion and magnification-bias effects,
resulting in different source redshift distributions $N(z)$.
In contrast to the former effect, the latter does not require source galaxies to be spatially 
resolved, but does require a stringent flux limit against incompleteness effects \citep{Umetsu+2011}.}
The mean lensing depth for a given population ($\alpha$) is given as
\begin{equation}
\beta_{(\alpha)}=\left[
\int_0^\infty\!dz\,\beta(z)N_{(\alpha)}(z) \right]
\left[
\int_0^\infty\!dz\,N_{(\alpha)}(z)
\right]^{-1}.
\end{equation}
It is useful to introduce the relative lensing strength of a source population 
with respect to a fiducial source at infinite redshift as
\citep{2001PhR...340..291B}.
\begin{equation}
w_{(\alpha)} = \beta_{(\alpha)} / \beta(z\to \infty).
\end{equation}
Then, the convergence and shear for such a fiducial reference source are  expressed as
$\kappa_{(\alpha)}(\btheta)=w_{(\alpha)}\kappa_{\infty}(\btheta)$ and 
$\gamma_{(\alpha)}(\btheta)=w_{(\alpha)}\gamma_\infty(\btheta)$, respectively.

\section{Cluster Lensing Methodology}
\label{sec:method}

In this section, we generalize the non-parametric Bayesian approach
of \citet{Umetsu+2011} for constructing projected mass profiles of individual clusters
from multi-probe gravitational lensing observations.
The Bayesian approach enables a full parameter-space extraction of model and
calibration parameters.
It is of particular importance to explore the entire
parameter space and investigate the parameter degeneracies,
arising in part from the mass-sheet degeneracy.

\subsection{Tangential Distortion}
\label{subsec:gt}
 
We construct radial profiles of the tangential distortion $g_+$ and 
the $45^\circ$-rotated component $g_\times$ as a function of
clustocentric radius
\citep[see][]{UB2008,Okabe+2010_WL}.
In the weak-lensing limit $g\approx \gamma$,
the azimuthally-averaged radial distortion profiles
satisfy the following {\it identity}
\citep{Kaiser1995}:
\begin{equation} 
\label{eq:loop}
\gamma_+(\theta) = \overline{\kappa}(<\theta)-\kappa(\theta); \ \ \ \gamma_\times(\theta)=0
\end{equation}
with $\kappa(\theta)$ the azimuthal average of $\kappa$ in the circular annulus of radius $\theta$
and $\overline{\kappa}(<\theta)$ the average convergence within a circular aperture of radius $\theta$,
defined as
$\overline{\kappa}(<\theta)=(\pi\theta^2)^{-1}\int_{\small |\btheta'|\le\theta}\! d^2\theta'\,\kappa(\btheta')$.

With the assumption of quasi-circular symmetry in the
projected cluster mass distribution $\kappa(\btheta)$ \citep[see][]{UB2008},  
the azimuthally-averaged tangential distortion is related to $\kappa$ as
\begin{equation}
\label{eq:gt}
g_+(\theta; w_g) =
\frac{\overline{\kappa}(<\theta)-\kappa(\theta)}{1-\kappa(\theta)}
=\frac{w_g[\bar{\kappa}_\infty(<\theta)-\kappa_\infty(\theta)]}
{1-w_g\kappa_\infty(\theta)},
\end{equation}
where $w_g$ is the population-averaged relative lensing strength (see Section \ref{subsec:pop}).
In the absence of higher order effects, weak
lensing induces only curl-free tangential distortions (Equation (\ref{eq:loop}). 
In practice, the presence of $\times$ modes can thus be used to check for systematic errors.

\subsection{Magnification Bias}
\label{subsec:magbias}

The source number counts for a given magnitude cutoff $m_{\rm cut}$
are modified in the presence of lensing as \citep{1995ApJ...438...49B}
\begin{equation} 
\label{eq:magbias}
N_\mu(<m_{\rm cut}) = N_0(<m_{\rm cut})\mu^{2.5s-1}(\btheta) \equiv N_0 b_\mu(\btheta;s), 
\end{equation}   
where $N_0(<m_{\rm cut})$ is the unlensed source counts,
approximated locally as a power-law with slope $s=d\log_{10} N_0(<m_{\rm cut})/dm_{\rm cut} >0$.
In the weak-lensing limit, $b_\mu(\btheta;s) \approx (5s-2)\kappa(\btheta)$, so that the 
signal-to-noise ratio (S/N) scales as $N_0^{1/2}(m_{\rm cut}) |5s(m_{\rm cut})-2|$
\citep{Rozo+Schmidt2010}.
For a maximally-depleted population of sources with $s=0$, $b_\mu(\theta)=\mu^{-1}(\theta)$.
The net magnification effect on the source counts vanishes ($b_\mu=1$) when $s=0.4$.

For a mass profile analysis, we calculate the surface number
density $n_\mu(\theta)=dN_\mu(\theta)/d\Omega$
of background sources as a function of clustocentric radius, by azimuthally
averaging $N_\mu(\btheta)$ \citep[for details, see][]{Umetsu+2011}.
The magnification bias is then expressed as
 $n_\mu(\theta)= n_0\mu^{2.5s-1}(\theta)$ with $n_0=dN_0(<m_{\rm cut})/d\Omega$ 
 and $\mu$ the magnification,\footnote{
Since magnification is nonlinear with respect to $\kappa_\infty$,
Equation (\ref{eq:mu}) is, strictly speaking, only valid for circularly symmetric lenses, or 
applicable in the weak-lensing regime where magnification is linearly
related to $\kappa_\infty$.
In our multi-probe approach, the number of data constraints can be sufficiently large
to break degeneracies in parameter space. In such a case, one can exclude
from the joint analysis magnification constraints in the nonlinear regime, and check for the presence of systematics.
}
\begin{equation}
\label{eq:mu}
\mu(\theta;w_\mu) = \frac{1}{|[1-w_\mu\kappa_\infty(\theta)]^2-w_\mu^2\gamma_{+,\infty}^2(\theta)|}.
\end{equation}
From Equations (\ref{eq:loop}) and (\ref{eq:mu}), the magnification
profile for a given source population
can be uniquely specified by the projected mass density profile
$\kappa_\infty(\theta)$.  

\subsection{Color-dependent Magnification Bias}
\label{subsec:relmagbias}

Deep multi-band photometry spanning a wide wavelength range allows us to identify distinct
populations of background galaxies in object color-color space (e.g., $B-R_{\rm c}$ versus $R_{\rm c}-z'$), as
demonstrated by recent cluster weak-lensing work based on Subaru 
observations \citep[e.g.,][]{Medezinski+2010, Medezinski+2011,
Umetsu+2010_CL0024, Umetsu+2011, Umetsu+2012}.
Since a given flux limit corresponds to different intrinsic luminosities
at different source redshifts, 
source-count measurements of distinctly-different background populations 
probe different regimes of magnification-bias effects.

The bias is strongly negative for red background galaxies at 
$\langle z\rangle \sim 1.1$ 
with a flat faint-end slope $s\sim 0.1$,
resulting in a net count depletion \citep{UB2008,Umetsu+2011,Umetsu+2012},
as dominated by the geometric area distortion:
$b_\mu\approx \mu^{-0.75}$.
For a depleted source population with $s\ll 0.4$,
the S/N increases progressively as $m_{\rm cut}$ increases.

On the other hand, the faint blue population of background galaxies,
lying at $\langle z\rangle \sim 2$
\citep{Lilly+2007,Medezinski+2010,Umetsu+2010_CL0024,Umetsu+2011,Umetsu+2012},  
tends to have a steep intrinsic slope 
close to the lensing invariant one ($s=0.4$).
For such a population,
source selection can be optimized to maximize the overall S/N \citep{Rozo+Schmidt2010},
by choosing a brighter $m_{\rm cut}$ (i.e., at a lower $n_0$) 
corresponding to a steeper count slope, say $s\sim 0.5$,
so that the bias is mildly positive and a net density enhancement results:
$b_\mu\approx \mu^{0.25}$.

Hence, combining the distinct blue and red background populations probes
a wider range of levels of magnification bias \citep{Broadhurst1995}. 
The relative magnification bias, defined as the ratio of the blue to red
galaxy counts, scales as
$b_\mu(\btheta; {\rm blue})/b_\mu(\btheta; {\rm red})
\approx \mu(\btheta)^{2.5\left[s({\rm blue})-s({\rm red})\right]}$,
For the count slopes quoted above,
$s({\rm blue})\approx 0.5$ and $s({\rm red})=0.1$, 
so that $b_\mu({\rm blue})/b_\mu({\rm red})\approx \mu$.

In general, combining independent count measurements 
from multiple source  populations has two major advantages to the
cluster lensing analysis:  First, it improves the
statistical precision of cluster lensing measurements.
Second, it has the further advantage of reduced sensitivity to the intrinsic angular
clustering of source galaxies, because distinctly different source
populations are spatially uncorrelated \citep{Broadhurst1995}.
In the linear weak-lensing regime (galaxy-galaxy and galaxy-group lensing),
the optimal weighting of magnification signals from different source
populations with different count slopes has been explored 
by several authors
\citep{Menard+Bartelmann2002, Scranton+2005, Hildebrandt+vanWaerbeke+Erben2009}. 

\subsection{Nonlinear Effect on the Source-averaged Weak-lensing Fields}
\label{subsec:nlin}   

In general, the weak-lensing effects in the cluster regime
are nonlinearly related to the underlying lensing potential.
Hence, in general, 
the averaging operator with respect to the source redshift
distribution $N(z)$ acts nonlinearly on the redshift-dependent components
in the cluster lensing observables. 
Our methodology here can be formally generalized to take
full account of the nonlinear effect on the source-averaged lensing profiles,
by replacing $g_+(w_g)$ and $b_\mu(w_\mu)$ 
in Equations (\ref{eq:gt}), (\ref{eq:magbias}),  and 
(\ref{eq:mu}) as \citep{Seitz1997}
\begin{eqnarray}
g_+(w_g) &\to&   \left[\int_0^\infty\!dz\,g_+(w[z])N_g(z) \right] 
\left[\int_0^\infty\!dz\,N_g(z) \right]^{-1}\nonumber\\
  &\equiv& \langle g_+\rangle\\
b_\mu(w_\mu)&\to& \left[\int_0^\infty\!dz\,b_\mu(w[z])N_\mu(z) \right] 
\left[\int_0^\infty\!dz\,N_\mu(z) \right]^{-1}\nonumber\\
 &\equiv& \langle b_\mu\rangle,
\end{eqnarray} 
where $N_g(z)$ and $N_\mu(z)$ are the respective redshift distribution
functions for the shape-distortion and magnification-bias measurements.

In the mildly-nonlinear regime, it is often sufficient 
to apply a low-order approximation using low-order moments of the
source-averaged lensing depth, 
neglecting higher-order correction terms.
For details, see Appendix \ref{appendix:nonlin}.
In particular, when 
the characteristic mean redshift $\langle z_s\rangle$
of source galaxies is sufficiently high compared to the cluster redshift
$z_l$ \citep[see][]{2001PhR...340..291B}, 
we can safely assume that all sources are at the same effective redshift,
corresponding to the mean lensing depth of the source
population
\citep[e.g.,][]{Medezinski+2010,Medezinski+2011,Umetsu+2010_CL0024,Umetsu+2011,Umetsu+2012}.

\subsection{Strong Lensing}
\label{subsec:sl}

The Einstein radius $\theta_{\rm E}$ is a characteristic size-scale of
strong lensing, tightly related to the cylinder mass it encloses,
$M(<\theta_{\rm E})$
\citep{1996astro.ph..6001N,Broadhurst+Barkana2008,Oguri+Blandford2009,Oguri+2012_SGAS}.
It describes the area enclosed by the tangential critical curve,
within which multiple imaging can occur due to the high surface mass
density of the lens.
For an axisymmetric lens, the average mass density within this critical area
is equal to $\Sigma_{\rm crit}$,
thus enabling us to directly estimate the enclosed mass
by $M(<\theta_{\rm E})=\pi(D_l\theta_{\rm E})^2\Sigma_{\rm crit}$.

In general, 
detailed strong-lens modeling with many sets of multiple images
allows us to determine the critical curves
with great accuracy, which then provides robust estimates of the projected mass 
enclosed by the critical area with an effective Einstein radius
$\theta_{\rm E}$ \citep[e.g.,][]{Zitrin+2011_A383}.
Accordingly, the enclosed mass profile
\begin{equation}
\label{eq:m2d}
M(<\theta)=\Sigma_{\rm crit}
\int_{\small|\btheta'|<\theta}\kappa(\btheta')\,d^2\theta'
=\pi (D_l\theta)^2\Sigma_{{\rm crit},\infty}\overline{\kappa}_\infty(<\theta)
\end{equation}
 at the
location around $\theta_{\rm E}$ is 
less sensitive to modeling assumptions and approaches 
\citep[see][]{Jullo+2007_Lenstool,Coe+2010,Umetsu+2012,Oguri+2012_SGAS},
serving as a fundamental observable quantity in the strong-lensing
regime
\citep[][Section 6.1]{Coe+2010}.
On the other hand, the radial profile slope of $\Sigma(\theta)$ is ill constrained 
owing to the mass-sheet (steepness) degeneracy (see Equation (\ref{eq:invtrans})).
Since the size of the Einstein ring $\theta_{\rm E}(z)$ grows with the source redshift $z$,
multiple sets of multiple images spanning
a wide range of source redshifts can be used to construct a reliable inner mass profile
$M(<\theta)$
 \citep[e.g.,][]{2005ApJ...621...53B,Saha+Read2009,Zitrin+2009_CL0024}.

\subsection{Bayesian Cluster Mass-profile Reconstruction}
\label{subsec:bayesian}

\subsubsection{Multi-probe Cluster Lensing Constraints}


We consider all possible lensing information available
in the cluster regime, 
namely enclosed mass estimates $M(<\theta)$ from strong lensing, 
tangential lens distortion $g_+(\theta)$ and magnification bias $n_{\mu}(\theta)$
measurements for multiple independent populations of background galaxies:
\begin{equation}
\label{eq:observable}
\{M_i\}_{i=1}^{N_{\rm sl}}, \{g_{+,i}\}_{i=1}^{N_{\rm wl}},
 \{n_{\mu(\alpha),i}\}_{i=1}^{N_{\rm wl}}\ \ (\alpha=1,...,N_{\rm col})
\end{equation}
with 
$N_{\rm col}$ the number of color-selected background samples for source-count measurements.
We measure the lens distortion and count profiles in the subcritical lensing regime (i.e., outside the critical curves, $\theta>\theta_{\rm E}$),
using the same grid of clustocentric annuli ($i=1,2,...,N_{\rm wl}$).
Hence, there are a total of $N_{\rm tot}\equiv (1+N_{\rm col})N_{\rm wl} + N_{\rm sl}$ lensing 
constraints including $N_{\rm sl}$ central projected mass estimates from strong lensing.

\subsubsection{Joint Likelihood Function}
\label{subsubsec:likelihood}

We construct a discretized mass profile from multi-probe cluster lensing 
constraints, 
extending the Bayesian approach by \citet{Umetsu+2011}.
In the Bayesian framework, we sample from the posterior probability
density function (PDF) of the underlying signal  $\bs$ given the data
$\bd$, $P(\bs|\bd)$.  
Expectation values of any statistic of $\bs$ shall converge
to the expectation values of the a posteriori marginalized PDF,
$P(\bs|\bd)$.  The covariance matrix $C$ of $\bs$ is obtained from the
resulting posterior sample. 
With the covariance matrix $C$, 
we introduce an estimator for the signal-to-noise ratio (S/N)
for detection of $\bs$,
integrated over the radial range considered \citep{UB2008}:
\begin{equation}
\label{eq:sn}
{\rm (S/N)^2} = \displaystyle{\sum}_{i,j} s_i {C}^{-1}_{ij} s_j
= \bs^{t} C^{-1} \bs.
\end{equation}

In our problem,
the signal $\bs$ is a vector containing 
the binned convergence profile $\{\kappa_{\infty,i}\}_{i=1}^N$
with $N=N_{\rm wl}+N_{\rm sl}$
and the average
convergence within the innermost aperture radius $\theta_{\rm min}$ 
for strong-lensing mass estimates
$\overline{\kappa}_{\infty,{\rm min}}\equiv
\overline{\kappa}(<\theta_{\rm min})$,\footnote{If there is no
strong-lensing constraint available ($N_{\rm sl}=0$),  
$\overline{\kappa}_{\rm min}$ represents the average convergence within
the inner radial boundary of weak-lensing observations. 
See \citet{Umetsu+2011}.} 
so that 
\begin{equation}
\bs=\{ \overline{\kappa}_{\infty, {\rm min}}, \kappa_{\infty,i}\}_{i=1}^N
\end{equation}
specified by $(N+1)$ parameters. 

The Bayes' theorem states that
\begin{equation}
P(\bs|\bd) \propto P(\bs) P(\bd|\bs),
\end{equation}
where ${\cal L}(\bs)\equiv P(\bd|\bs)$ is the 
likelihood of the data
given the model ($\bs$), and $P(\bs)$ is the prior PDF for the model parameters.
The joint ${\cal L}(\bs)$ function for
multi-probe cluster lensing observations is given as a product of 
their separate likelihoods, 
\begin{equation}
{\cal L}={\cal L}_{\rm wl}{\cal L}_{\rm sl}={\cal
 L}_g{\cal  L}_\mu {\cal L}_{\rm sl},
\end{equation}
where ${\cal L}_{\rm wl}={\cal L}_g{\cal L}_\mu$ and ${\cal L}_{\rm sl}$
are the likelihood functions for weak and strong lensing, respectively,
defined as 
\begin{eqnarray}
\ln{\cal L}_{g}(\bs)&=& -\frac{1}{2}\sum_{i=1}^{N_{\rm
 wl}}\frac{[g_{+,i}-\hat{g}_{+,i}(\bs;w_g)]^2}{\sigma^2_{+,i}},\\
\ln{\cal L}_\mu(\bs)&=& -\frac{1}{2}\sum_{\alpha=1}^{N_{\rm col}}\sum_{i=1}^{N_{\rm wl}}
\frac{[n_{\mu(\alpha),i}-\hat{n}_{\mu(\alpha),i}(\bs;w_{(\alpha)})]^2}{\sigma^2_{\mu(\alpha),i}},\\
\label{eq:Lsl}
\ln{\cal L}_{\rm sl}(\bs)&=&-\frac{1}{2} \sum_{i=1}^{N_{\rm
 sl}}\frac{[M_{i}-\hat{M}_{i}(\bs)]^2}{\sigma^2_{M,i}},
\end{eqnarray}
where $(\hat{g}_{+}, \hat{n}_{\mu(\alpha)}, \hat{M})$ are the
theoretical predictions for the corresponding observations,
and all these profiles can be uniquely specified by $\bs$ (see Appendix
\ref{appendix:estimators}).  

The errors $\sigma_{+,i}$ for $g_{+,i}$ due primarily to the variance of the 
intrinsic source ellipticities can be conservatively
estimated from the data 
using bootstrap techniques. The errors $\sigma_{\mu(\alpha),i}$ for
$n_{\mu(\alpha),i}$ 
include both contributions from Poisson counting errors 
and contamination due to intrinsic clustering of each source population 
\citep{Umetsu+2011}. 
The strong-lensing mass estimates $M_i$ and errors $\sigma_{M,i}$ 
can be derived from detailed modeling of 
multiply-lensed images (see Section
\ref{subsec:sl}).  The inner mass profile $M_i$ can be measured at
several independent aperture radii when multiple sets of multiple images
are available at various source redshifts.\footnote{Each set of
multiply lensed images constrains the mass enclosed within
their radii, $M(<\theta)$.}   The covariance between binned
$\kappa$ values naturally arises because they are to satisfy
the observed cumulative mass constraints \citep{Coe+2010}.

\subsubsection{Priors}
\label{subsubsec:prior}
 
For each parameter of the model $\bs$, we consider a flat uninformative
prior with a lower bound of $\bs=0$, that is,
$\overline{\kappa}_{\infty,{\rm min}}>0$ and $\kappa_{\infty,i} >0$.
Additionally, we account for the calibration uncertainty in the
observational parameters, such as 
the relative lensing strength $w_{(\alpha)}$,
the count normalization and slope parameters
($n_{0(\alpha)},s_{(\alpha)}$) 
for each color sample \citep[see][]{Umetsu+2011}:
\begin{equation}
\bc=\{w_g,w_{(\alpha)}, n_{0(\alpha)}, s_{(\alpha)}\}_{\alpha=1}^{N_{\rm col}},
\end{equation}
giving a set of $(3N_{\rm col}+1)$ calibration parameters to marginalize over.

In practice, the count normalization and slope parameters can be estimated
from the  counts in cluster outskirts using wide-field imaging
data.  The mean depth of background samples can be either measured 
from well-calibrated photometric redshifts \citep{Umetsu+2012}, or
estimated from deep multi-band photometry, such as the 30-band COSMOS
database \citep{Ilbert+2009_COSMOS}.

\subsection{Implementation}
\label{subsec:implementation}

We implement our method using a Markov Chain Monte Carlo (MCMC) 
approach with Metropolis-Hastings sampling, by following the
prescription outlined in \citet{Umetsu+2011}. 
For Bayesian parameter estimation, we use the location of the marginal maximum a
posteriori probability (MMAP) for each model parameter, using the
bisection method in conjunction with bootstrap techniques
\citep{Umetsu+2011}. 
The covariance matrix $C$ 
for the discrete mass profile $\bs$ is estimated from
the MCMC samples.
The method has been tested against synthetic shear+magnification catalogs
from simulations of analytical NFW lenses performed using the public package {\it glafic} \citep{Oguri2010_glafic}.
The results suggest that both maximum-likelihood (hereafter, ML) and
Bayesian MMAP solutions produce reliable reconstructions with unbiased profile 
measurements, 
so that this multi-probe lensing method is not sensitive to the
choice and form of priors,
when the shear and magnification are combined and hence
the mass-sheet degeneracy is fully broken.

\section{Applications to Cluster Lensing Observations: MACS1206}
\label{sec:app}

In this section, we apply our new method to the recent CLASH 
observations of MACS1206 presented by \citet{Umetsu+2012}, 
who performed a joint analysis of weak-lensing distortion and {\it
negative} magnification-bias measurements for a reconstruction of the 
projected cluster mass profile, then compared and combined with the
inner mass profile derived independently from their strong-lensing
analyses.  
Here we conduct a joint likelihood analysis of the full lensing constraints
from strong-lensing,  
weak-lensing distortion, positive and negative magnification-bias measurements,
demonstrating how combining multi-probe lensing constraints can improve
the mass profile reconstruction.

\subsection{MACS1206}
\label{subsec:m1206}

MACS1206 is an X-ray selected CLASH cluster \citep{Postman+2012_CLASH}
with a fairly relaxed appearance in optical, X-ray, and
Sunyaev-Zel'dovich effect (SZE) images, as well as in morphology of its
brightest cluster galaxy (BCG)
\citep{Ebeling+2009_M1206,Gilmour+2009,Umetsu+2012}. 
No significant offset is observed between the BCG, X-ray peak, and 
DM center of mass determined from detailed strong-lens modeling
\citep{Zitrin+2012_M1206,Umetsu+2012}.
For the cluster, a good agreement is obtained between the lensing, X-ray,
and SZE mass estimates \citep{Umetsu+2012,Rozo+2012_closing}, 
with a virial mass of $M_{\rm vir}=(1.1\pm 0.2)\times 10^{15}M_\odot\,h^{-1}$
\citep{Umetsu+2012},
indicating
the hot gas is not far from a state of hydrostatic equilibrium in
cluster potential well. 
On large scales, the cluster is embedded in elongated rich large-scale structures
as revealed by the galaxy and weak-lensing mass maps
\citep{Umetsu+2012}.

The cluster was observed deeply in 16 filters ranging from the UV through
the optical to the IR on the {\it Hubble Space Telescope}
(hereafter, {\it HST}) and in five optical passbands, $BVR_{\rm c}I_{\rm
c}z'$, with the wide-field Suprime-Cam \citep[$34\arcmin\times
27\arcmin$;][]{2002PASJ...54..833M} on 
the 8.2\,m Subaru telescope.   
The majority of Subaru observations were taken as part of the
Weighing the Giants project \citep{WtGI}.
We refer the reader to \citet{Zitrin+2012_M1206} and \citet{Umetsu+2012}
for details of the {\it HST} (strong lensing) and Subaru (weak
lensing) observations, respectively. 
We define the cluster center to be the location of the BCG, following
\citet{Umetsu+2012}. 

\subsection{Weak-lensing Shear and Magnification Constraints}
\label{subsec:clash-wl}


We derive radial profiles of 
lens distortion and magnification-bias from a
reanalysis of Subaru observations of \citet{Umetsu+2012}.
We have two color samples ($N_{\rm col}=2$) of blue and red
background populations for magnification measurements, and a full
background sample of blue+red galaxies for shape distortion
measurements, as defined by \citet[][see their Figure 3, Tables 3 and
4]{Umetsu+2012} using the background-selection method of
\citet{Medezinski+2010}.

For the red counts to measure the negative bias,
we have 13252 galaxies with a mean depth of 
$\beta({\rm red})=0.51\pm 0.02$,
at the magnitude limit of $m_{\rm cut}(z')=24.6$\,mag \citep{Umetsu+2012}.
For this the normalization and slope parameters are estimated as 
$n_0({\rm red})=11.4\pm 0.3$\,arcmin$^{-2}$ and $s({\rm red})=0.133\pm 0.047$
from the coverage-corrected source counts
in the outer region, $\theta\simgt 10\arcmin$.
For the blue counts to measure the positive bias,
we find that the central density enhancement decreases with increasing 
$m_{\rm cut}(z')$, vanishing at a deeper cut of 
$m_{\rm cut}(z')\sim 25$\,mag,
where the count slope is close to the lensing-invariant one, $s({\rm
blue})\simeq 0.4$.
Hence, the magnitude cutoff $m_{\rm cut}(z')$ has been chosen
to optimize the total S/N integrated over all radial bins
of the magnification signal (Section~\ref{subsec:magbias}); for this, we
find 2740 galaxies with
$\beta({\rm blue})=0.62\pm 0.06$,
$s({\rm blue})=0.532\pm 0.105$, 
and
$n_0({\rm blue})=1.98\pm 0.13$\,arcmin$^{-2}$
at $m_{\rm cut}(z')=24.0$\,mag.
Finally, we have 13123 background galaxies  with $\beta({\rm
back})=0.54\pm 0.03$ for the distortion measurements \citep{Umetsu+2012}.

We calculate the respective weak-lensing profiles in $N_{\rm wl}=10$
discrete radial bins from the cluster center, 
spanning the range $\theta=[0.6\arcmin,16\arcmin]$ with a constant
logarithmic radial spacing $\Delta\ln\theta\simeq 0.328$.
Combining the tangential distortion, blue and red count measurements, we
have a total of $30$ constraints from Subaru weak-lensing observations.

\begin{figure}[!htb] 
 \begin{center}
 \includegraphics[width=180mm,angle=0]{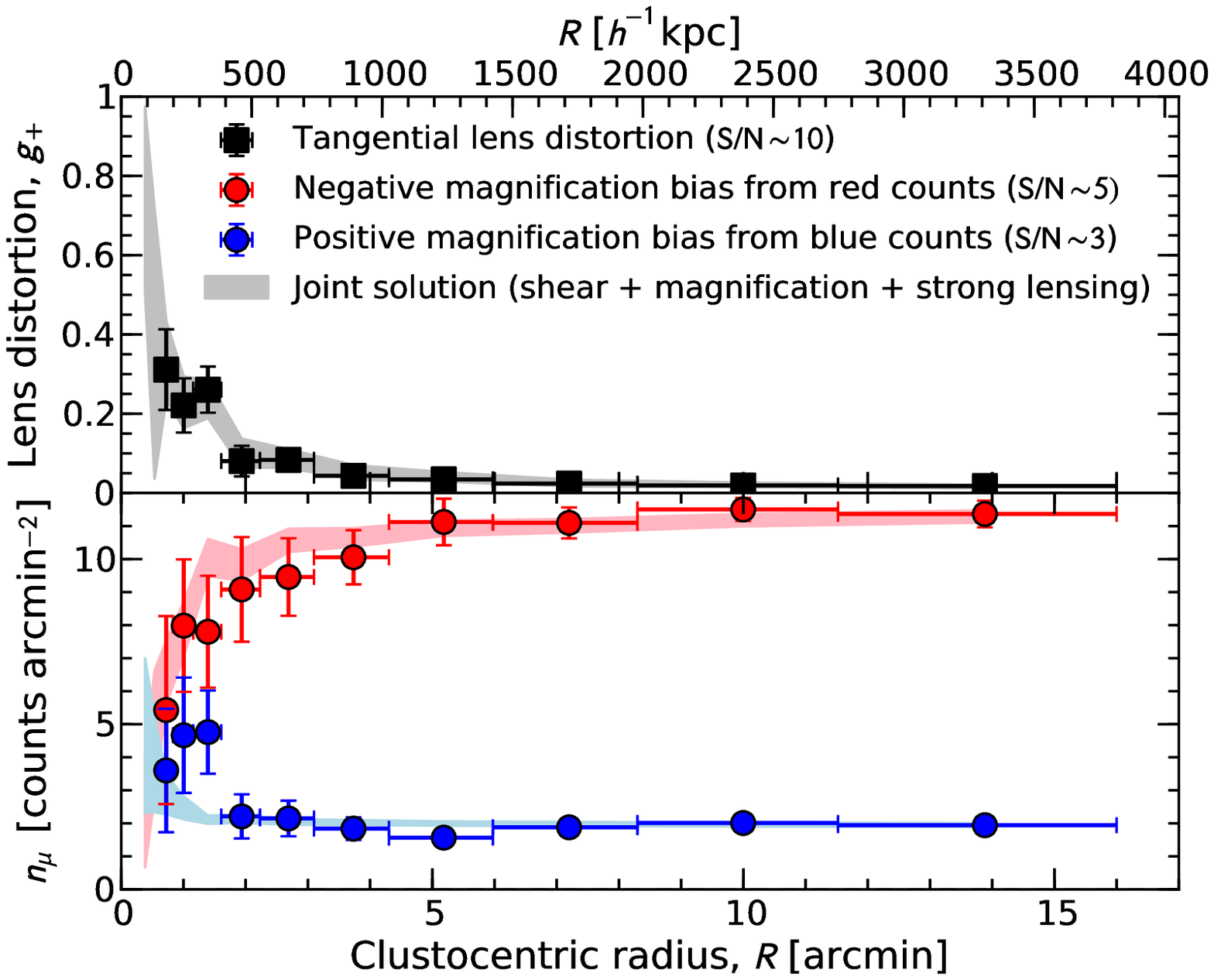} 
 \end{center}
\caption{
Cluster weak-lensing radial profiles of MACS1206 ($z_l=0.439$) obtained from a
 reanalysis of the Subaru $BR_{\rm c}z'$ data presented in \citet[][see
 their Figures 4 and 5]{Umetsu+2012}.
The top panel shows the tangential
reduced shear profile $g_+$ (squares) based on the full background
 sample. 
The bottom panel shows the  
coverage-corrected number-count profiles $n_\mu$ for
 flux-limited samples of blue and red background galaxies (circles).
The error bars include contributions from
Poisson counting uncertainties and contamination due to intrinsic angular
 clustering of each source population.
For the red sample, a strong radial depletion of the source counts
 is seen toward the cluster center due primarily to magnification of the 
 sky area,  
while a slight enhancement of blue counts 
 is present in the innermost radial bins due to the effect of positive magnification bias. 
Also shown for each observed profile is the joint Bayesian
 reconstruction from combined strong-lensing,
 weak-lensing tangential distortion, positive and negative
 magnification-bias 
 measurements (see also Figure \ref{fig:mass}).
\label{fig:wldata}
}
\end{figure}

In Figure \ref{fig:wldata} we show the resulting 
lens-distortion (black),
positive (blue) and negative (red)
magnification-bias measurements of MACS1206 as a function of
clustocentric radius.
We find an integrated S/N of 10.2, 2.9, and 4.7 for the above respective
measurements (defined in analogy to Equation (\ref{eq:sn})). 
The total S/N for the combined weak-lensing
measurements is estimated as ${\rm (S/N)}_{\rm
WL}=\sqrt{10.2^2+2.9^2+4.7^2}\simeq 11.6$.

\subsection{Strong-lensing Constraints}
\label{subsec:clash-sl}

The central mass distribution of MACS1206 has been tightly constrained by detailed
strong-lensing analyses based on CLASH {\it HST} imaging
and Very Large Telescope/VIMOS spectroscopic observations
\citep{Zitrin+2012_M1206,Umetsu+2012}.
There are a total of 50 multiply-lensed
images of 13 background sources identified for this cluster
\citep{Zitrin+2012_M1206}, spanning  
a wide range of source redshifts, $1\simlt z_s\simlt 5.5$, spread fairly evenly
over the central region, $3\arcsec\simlt \theta\simlt
1\arcmin$.  

For a source at $z_s=2.54$, the
tangential critical curve encloses an area with an effective Einstein
radius of $\theta_{\rm E}=28\arcsec \pm 3\arcsec$; 
for another lower-redshift system with
$z_s=1.03$ \citep{Ebeling+2009_M1206}, the effective Einstein radius of the critical area is
$\theta_{\rm E}=17\arcsec\pm 2\arcsec$ \citep[see][]{Zitrin+2012_M1206,Umetsu+2012},\footnote{\citet{Zitrin+2012_M1206} used the position and redshift of 32 secure multiple images of nine 
systems to constrain their mass model.} corresponding to
model-independent projected mass estimates of 
$M(<17\arcsec)=5.80^{+1.28}_{-1.44}\times 10^{13}M_\odot\,h^{-1}$
and
$M(<28\arcsec)=1.11^{+0.22}_{-0.25}\times 10^{14}M_\odot\,h^{-1}$:
These are shown to be in broad agreement with the aperture mass
measurements (Equation (\ref{eq:m2d}))
obtained from several complementary strong-lensing analyses via a
variety of modeling methods and approaches \citep[for details,
see][their Figure 7]{Umetsu+2012}.  

In the present analysis, we use the double Einstein-radius
constraints 
on the  projected total mass $M(<\theta)$ derived at two distinct source
redshifts,  $z_s=1.03$ and $2.54$ ($N_{\rm sl}=2$).
The total S/N for the combined strong-lensing constraints is
obtained as $({\rm S/N})_{\rm SL}\simeq 6.3$.


\subsection{Results}
\label{subsec:results}

Here we construct discrete mass profiles of MACS1206 from our
multi-probe cluster lensing observations, demonstrating how 
additional strong-lensing and positive magnification-bias information
can improve the mass profile reconstruction.
For our full-lensing analysis,
we have a total of  $N_{\rm tot}=(1+N_{\rm col})N_{\rm wl}+N_{\rm sl}=32$
constraints from strong-lensing ($M$), weak-lensing distortion ($g_+$),
and positive and negative magnification-bias measurements ($n_{\mu}({\rm
blue}), n_{\mu}({\rm red})$), in the range
$\theta_{\rm min}=\theta_{\rm E}(z_s=1.03)=17\arcsec\le \theta \le 16\arcmin = \theta_{\rm max}$,
with $N_{\rm col}=2$, $N_{\rm wl}=10$, $N_{\rm sl}=2$, and $N=N_{\rm wl}+N_{\rm sl}=12$.  
The mass-profile model $\bs=\{ \overline{\kappa}_{\infty,{\rm
min}},\kappa_{\infty,i} \}_{i=1}^N$ is then 
described by $N+1=13$ parameters (see Section~\ref{subsubsec:likelihood}). 
Additionally, we have seven calibration parameters ($\bc$)
to marginalize over (Section~\ref{subsubsec:prior}). 
The projected cumulative mass profile $M(<\theta)$ is given by
integrating the density profile $\bs$ (see also Appendix
\ref{appendix:estimators}) as
\begin{eqnarray}
M(<\theta)&=&\pi (D_l\theta_{\rm min})^2 \Sigma_{{\rm crit},\infty} 
 \overline{\kappa}_{\infty,{\rm min}}\nonumber\\
&&+2\pi D_l^2 \Sigma_{{\rm crit},\infty}
\int_{\theta_{\rm min}}^{\theta}\! d\ln\theta'\,\theta'^2\kappa_\infty(\theta').
\end{eqnarray}

\begin{figure}[!htb] 
 \begin{center}
 \includegraphics[width=180mm,angle=0]{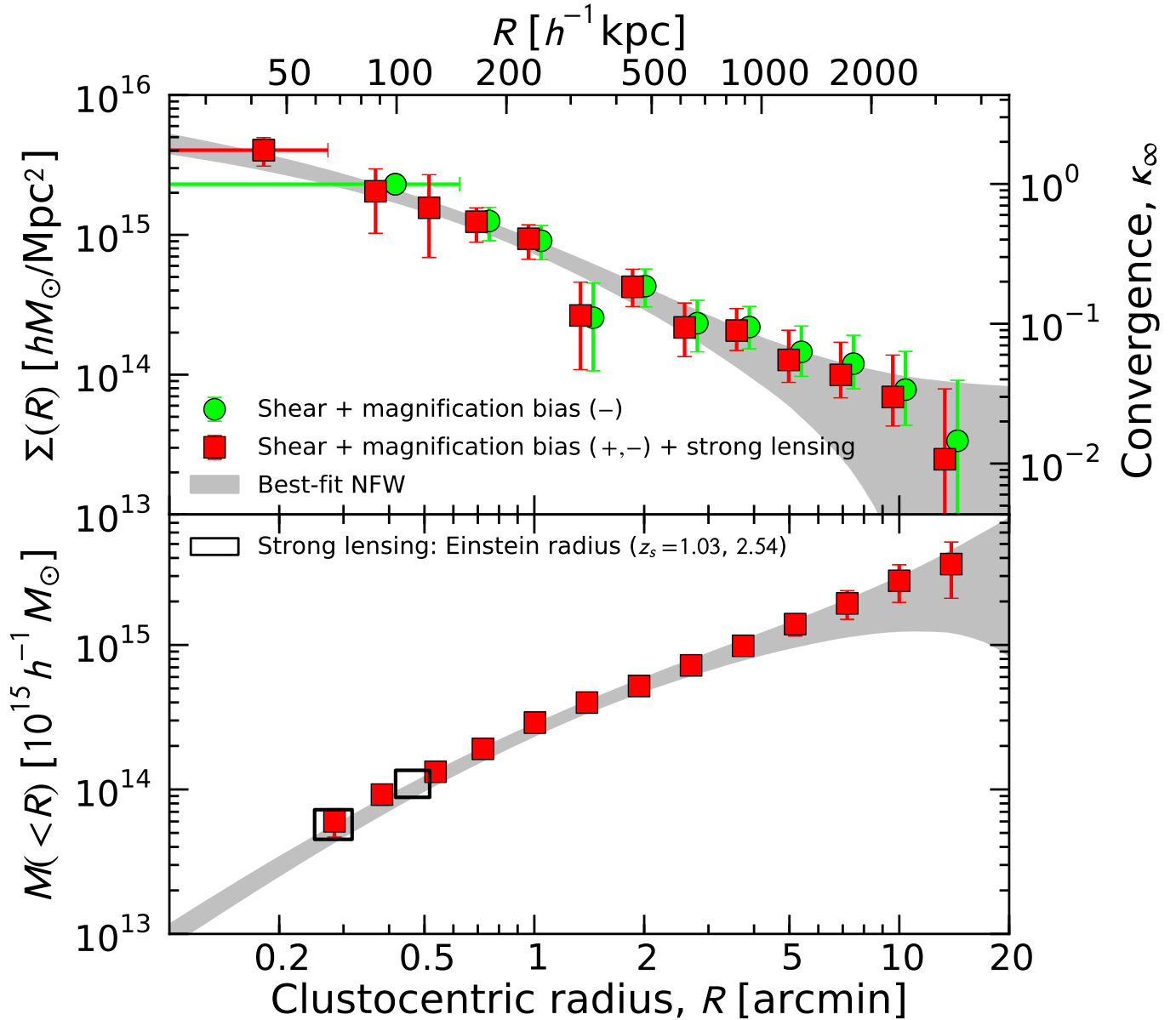} 
 \end{center}
\caption{
Effects of including additional strong-lensing and positive
 magnification-bias  constraints. 
The top panel shows the projected mass density profile $\Sigma(R)$ of MACS1206
 derived from multi-probe cluster lensing measurements.
For each case, the innermost central bin
$\overline{\Sigma}(<R_{\rm min})$ is marked with a
 horizontal bar.
The green circles show the $\Sigma(R)$ profile reconstructed
from a joint likelihood analysis of weak-lensing distortion and negative
 magnification-bias measurements (Figure \ref{fig:wldata}). 
The red squares are obtained from a joint analysis of 
the full lensing constraints, including additional strong-lensing (boxes,
 bottom) and positive magnification-bias (Figure \ref{fig:wldata})
 measurements.
The central weak-lensing bin $\overline{\Sigma}(<0.6\arcmin)$
is resolved into 3 radial bins when additional double Einstein-radius constraints
from strong lensing are included.
For visual clarity, the reconstructed mass profiles 
are horizontally shifted by $4\%$ with respect to each
 other.   
The red area in the bottom panel shows the corresponding enclosed mass profile $M(<R)$.
The boxes represent the double Einstein-radius constraints on the enclosed mass,
corresponding to background sources at $z_s=1.03$ and $2.54$ \citep[][see
 their Figure 7]{Umetsu+2012}.
In each panel, the gray area represents the best-fit NFW (+ mass-sheet) profile derived
from a joint analysis of the full-lensing constraints (red squares).
\label{fig:mass}
} 
\end{figure}

The results are shown in Figure \ref{fig:mass}.
First, we use only the weak-lensing distortion
and depletion constraints for our joint likelihood analysis
($\theta_{\rm min}=0.6\arcmin$, $\theta_{\rm max}=16\arcmin$;
$N_{\rm col}=1$, $N_{\rm tot}=20$, $N=N_{\rm wl}=10$),
corresponding to the weak-lensing analysis of \citet{Umetsu+2012}.\footnote{
In \citet{Umetsu+2012} the covariance matrix includes the contribution from the cosmic noise due to uncorrelated large-scale structure projected along the line of sight.}
From this, we find a consistent mass-profile solution $\bs$ (green circles)
with an integrated S/N of $12.1$.
The ML solution has a reduced $\chi^2_{\rm min}$ of
$4.33$ for 5 degrees of freedom (dof). 

Next, the mass-profile solution from our full-lensing constraints is
shown in Figure \ref{fig:mass} with red squares
($\chi^2_{\rm min}/{\rm dof}=13.3/12$). 
It is demonstrated that, when the strong-lensing information is included, the
central weak-lensing bin $\overline{\kappa}_{\infty}(<0.6\arcmin)$ is
resolved into $N_{\rm sl}+1=3$ radial bins, hence improving the determination of the
inner mass profile.
The S/N ratio in the recovered mass profile is 
$13.6$, which is consistent within uncertainties with the total S/N obtained from the
linearly-combined full-lensing constraints, 
$\sqrt{({\rm S/N})_{\rm WL}^2 + ({\rm
S/N})^2_{\rm SL}} = \sqrt{11.6^2+6.3^2}\simeq 13.2$.
We find here a minor statistical improvement from adding the blue source
counts, because the positive magnification-bias effect is only
marginally detected in our analysis ($\sim 3\sigma$).
In Figure \ref{fig:wldata} we also display the joint Bayesian
reconstruction of each observed radial profile, ensuring consistency and reliability
of our lensing analysis and methods.

\section{Summary and Discussion}
\label{sec:summary}

In this paper we have developed a non-parametric Bayesian method for
a direct reconstruction of the projected cluster mass profile from
a multi-probe cluster lensing analysis (Section~\ref{subsec:bayesian}),
combining independent 
strong-lensing, weak-lensing shear and magnification measurements.
This multi-probe approach improves the accuracy and precision of
the cluster lens reconstruction,  
effectively breaking the mass-sheet degeneracy (Section~\ref{subsec:sheet}). 
This work extends our earlier work by \citet{Umetsu+2011}
to include multiple populations of background
sources for magnification-bias measurements
(Sections~\ref{subsec:magbias} and \ref{subsec:relmagbias})  
and enclosed total mass estimates in the central
strong-lensing region (Section~\ref{subsec:sl}).  This flexible method
applies to both the strong- and weak-lensing regimes for full radial
coverage (Section~\ref{sec:app}).

Magnification bias depends on the intrinsic properties of the source
luminosity function as well as the source distance, in contrast to the
purely-geometric color-independent distortion effects.
The combination of count measurements for distinct blue and red 
background populations probes a wider range of levels of magnification bias
(Section~\ref{subsec:relmagbias}), 
boosting the significance of cluster lensing measurements
(Section~\ref{sec:app}). 
Furthermore, combining spatially-uncorrelated distinct populations of
background galaxies will help reduce the reconstruction bias due to
their intrinsic angular clustering
\citep{1995ApJ...438...49B,Broadhurst1995}. 
It is also interesting to note that, unlike the shearing effect,
magnification is in principle sensitive to the sheet-like structure
(Section~\ref{subsec:sheet}), so that making accurate magnification
measurements is crucial for a robust statistical detection  of the 
{\it two-halo term} contribution due to large-scale structure associated 
with the central clusters \citep{Oguri+Takada2011,Oguri+Hamana2011}.


Adding strong-lensing information to weak-lensing
is needed to provide tighter constraints on the inner density profile
\citep{2007ApJ...668..643L,UB2008,Oguri+2009_Subaru,Newman+2009_A611,Merten+2009,Oguri+2012_SGAS}.
The full mass profile constraints on individual clusters, derived from a joint
weak+strong lensing analysis, can be further stacked
together  to increase the statistical precision of the average mass profile
determination. 
Recently \citet{Umetsu+2011_stack} performed a comprehensive stacked lensing analysis
of four similar mass clusters \citep{Umetsu+2011},
by combining weak-lensing derived $\kappa(\theta)$ profiles
with their respective inner $\kappa(\theta)$ profiles derived independently from detailed
strong-lens modeling.  In contrast, the new method developed here
allows us to construct a joint weak+strong lensing likelihood
function of the underlying $\kappa$ signal,
by explicitly combining weak-lensing measurements with strong-lensing
enclosed mass estimates $M(<\theta)$, a more fundamental observable
than $\kappa(\theta)$ in the strong-lensing analysis \citep{Coe+2010}.
Therefore, our new approach permits a direct error propagation
and thus a more accurate derivation of the reconstruction covariance
matrix $C$. 

When combining strong and weak lensing constraints in different regimes of
signal strength and significance, it is crucial to 
account for possible systematic errors introduced by inherent modeling and prior assumptions,
especially for strong-lens modeling due to complex, nonlinear error propagation.
Recently, \citet{Umetsu+2012} introduced a regularization technique to 
obtain robust, conservative error estimates for the binned mass profile derived from strong lensing,
by calibrating the covariance matrix to eliminate very small eigenvalues associated with
large-scale modes where the constraints are weak and essentially driven by the prior.
It is straightforward to extend our joint likelihood analysis to include 
in Equation (\ref{eq:Lsl}) the covariance between radial bins.

     
An accurate determination of the cluster density profile for full radial
coverage 
is crucial for testing DM and alternative-gravity paradigms 
\citep{Newman+2012a,Newman+2012b,Narikawa+Yamamoto2012,Geller+2013,Silva+2013}. 
\citet{BEC2009} explored in detail  an extremely-light bosonic
dark-matter 
(ELBDM) model with a mass of the order $10^{-22}$\,eV, 
as an alternative to CDM to account for the perceived lack of small galaxies
relative to the $\Lambda$CDM model \citep{Klypin+1999,Peebles+Nusser2010}.
ELBDM with a de-Broglie wavelength of astronomical length scales, if it
exists, may well be in a ground-state Bose-Einstein condensate (BEC)
and hence described by a coherent wave function, behaving
effectively as a single scalar field \citep{Hu+2000}.
Intriguingly, \citet{BEC2009} showed that, 
ELBDM halos can form steepening density profiles of the form similar to
the standard CDM, irrespective
of whether halos form through accretion or merger.
However, during a collision between BECs interesting large-scale interference
occurs which will differ markedly from the behavior of 
cluster galaxies and standard collisionless CDM.  
Therefore, it is important to explore this class of DM
further via more extensive simulations for providing testable
predictions against detailed cluster lensing observations of both
relaxed and merging clusters.

The CLASH survey \citep{Postman+2012_CLASH} is particularly designed to
generate such useful lensing data, 
combining high-resolution 16-band {\it HST} imaging with
wide-field Subaru observations,
for a sizable sample of 20 X-ray selected ($T_X>5\,$keV)
relaxed clusters \citep{Zitrin+2011_A383,Coe+2012_A2261,Umetsu+2012},
free of lensing-selection bias,
and a lensing-selected sample of five high-magnification clusters,
the majority of which  are physically-interacting merging systems
\citep[][]{Zitrin+2011_MACS,Zitrin+2013_M0416}{Medezinski+2013}.
A stacked analysis of the X-ray selected CLASH clusters 
(with a halo bias factor of $b_h\sim 8$), combining
all lensing-related effects in the cluster regime, is highly desirable
for a definitive determination of the representative mass profile out to
beyond the virial radius, where the two-halo term is
expected to be detectable in the averaged $\kappa$ profile
\citep{Tinker+2010,Oguri+Hamana2011,Silva+2013}, 
providing a firm basis for a detailed comparison
with the standard $\Lambda$CDM paradigm and a wider examination of
alternative scenarios \citep{Gao+2012_Phoenix,Narikawa+Yamamoto2012,Silva+2013}.



\acknowledgments
I thank the anonymous referee for providing valuable comments
and constructive suggestions.
This work was made possible in part by the availability of high-quality
lensing data produced by the CLASH team.
I express my gratitude to
all members of the CLASH team who made the data analyzed here possible.
I thank Elinor Medezinski,
Mario Nonino, 
and Alberto Molino, 
for their valuable contributions to the weak-lensing data analysis.
I am very grateful to Marc Postman for his generous support and encouragement.
I acknowledge 
Nobuhiro Okabe,
Tom Broadhurst, 
Doron Lemze,
Adi Zitrin, 
Dan Coe,  
Julian Merten,
Sherry Suyu,
Margaret Geller,
and Tzihong Chiueh
for fruitful discussions and comments.
The work is partially supported by the National Science Council of Taiwan
under the grant NSC97-2112-M-001-020-MY3 and by 
the Academia Sinica Career Development Award.

\clearpage


\appendix

\section{Nonlinear Effect on the Source-averaged Lensing Fields}
\label{appendix:nonlin}

\subsection{Reduced Shear}

The reduced gravitational shear, $g=\gamma/(1-\kappa)$, is nonlinear in
$\kappa$, so that the averaging operator with respect to the source
redshift acts nonlinearly on $\kappa$.
In general, a wide spread of the redshift distribution of background
galaxies, in conjunction with the single source-plan approximation, may
lead to an overestimation of the gravitational shear in the nonlinear
regime \citep{2000ApJ...532...88H}.

Too see this effect, we expand the reduced shear $g=g(z)$ with respect
to $\kappa=\kappa(z)$ as
\begin{equation}
g=\gamma/(1-\kappa)=w\gamma_\infty(1-w\kappa_\infty)^{-1}
=w\gamma_\infty\sum_{k=0}^\infty\left(w\kappa_\infty\right)^k.
\end{equation}
Hence, the reduced shear averaged over the source redshift distribution
is expressed as
\begin{equation}
\langle g\rangle 
=\gamma_\infty\sum_{k=0}^\infty \langle w^{k+1}\rangle \kappa_\infty^k,
\end{equation}
where the angular brackets denote an ensemble average over the redshift
distribution $N(z)$ of background sources.
In the weak-lensing limit where $\kappa_\infty,|\gamma_\infty|\ll 1$,
$\langle g\rangle\approx \langle w\rangle\gamma_\infty
=\langle\gamma\rangle$.   The next order of approximation is given by 
\begin{equation}
\label{eq:s97}
\langle g\rangle \approx \gamma_\infty\left(
\langle w\rangle  + \langle w^2\rangle\kappa_\infty^2
\right)
\approx \frac{\langle w\rangle \gamma_\infty}{1-\kappa_\infty\langle
w^2\rangle/\langle w\rangle}.
\end{equation}
\citet{Seitz1997} found that Equation (\ref{eq:s97}) yields an excellent
approximation in the mildly-nonlinear regime of $\kappa_\infty\simlt
0.6$. 
Defining $f_w\equiv \langle w^2\rangle/\langle w\rangle^2$, 
a dimensionless quantity of the order of unity,
we have the
following expression for the source-averaged reduced shear valid in the
mildly-nonlinear regime:
\begin{equation}
\langle g\rangle
\approx \frac{\langle\gamma\rangle}{1-f_w\langle\kappa\rangle},
\end{equation}
with $\langle\kappa\rangle=\langle w\rangle\kappa_\infty$ and
$\langle\gamma\rangle=\langle w\rangle\gamma_\infty$ \citep{Seitz1997}. 
For clusters lying at relatively low redshifts, 
$\langle w^2\rangle\approx \langle w\rangle^2$
and 
$f_w\approx 1$, leading to the single source-plane approximation:
$\langle g\rangle\approx \langle \gamma\rangle/(1-\langle
\kappa\rangle)$. 
The level of bias introduced by this approximation is 
$\Delta g/g\approx (f_w-1)\langle \kappa\rangle$.
In typical ground-based deep observations of $z_l\simlt 0.5$
clusters
\citep{Okabe+2010_WL,Medezinski+2010,Umetsu+2010_CL0024,Umetsu+2011,Umetsu+2012},  
$\Delta f_w\equiv f_w-1$ is found to be of the order of several percent, so that the
relative error in the $g$ estimate is negligibly small in the
mildly-nonlinear regime.

\subsection{Magnification Bias}

First, let us consider a maximally-depleted sample of background sources
with $s=d\log_{10}N(<m)/dm=0$, for which the effect of magnification bias is purely
geometric, $b_\mu =\mu^{-1}$, and is insensitive to the intrinsic source
luminosity function.
In the nonlinear subcritical regime, the source-averaged
magnification bias is expressed as 
\begin{equation}
\langle b_\mu\rangle  =\langle \mu^{-1}\rangle
=1-2\langle \kappa\rangle- f_w\left(
|\langle\gamma\rangle|^2-\langle\kappa\rangle^2
\right)
=\mu^{-1}(\langle w\rangle)
+(f_w-1)\left(
\langle\kappa\rangle^2-|\langle\gamma\rangle|^2
\right),
\end{equation}
where $\mu^{-1}(\langle w\rangle)=(1-\langle
\kappa\rangle)^2-|\langle\gamma\rangle|^2$.
The error associated with  the single source-plane approximation is therefore
$\Delta b_\mu=(f_w-1)(\langle\kappa\rangle^2-|\langle\gamma\rangle|^2) = \Delta f_w(\langle\kappa\rangle^2-|\langle\gamma\rangle|^2)$, 
which is much smaller than unity for background source populations of
our concern in the mildly-nonlinear subcritical regime
($\langle\kappa\rangle\sim |\langle\gamma\rangle|\sim O(10^{-1})$).
It is therefore reasonable to use the single source-plane approximation
for calculating the magnification bias of depleted source populations
with $s \ll 0.4$.

In the positive regime of magnification bias ($s>0.4$),
on the other hand, interpreting the observed lensing signal (i.e., the
density enhancement) is more difficult, especially in the nonlinear
regime where the flux amplification factor is correspondingly large 
(say, $\mu\simgt 1.5$). 
Therefore, it requires detailed information about the intrinsic
source luminosity function to apply nonlinear corrections due to the
spread of the source redshift distribution.
In practice, the distant blue population of
background galaxies has a distinct, well-defined redshift distribution,
which is fairly symmetric and peaked at a mean redshift of $\langle
z_s\rangle \sim 2$
\citep{Lilly+2007,Medezinski+2010,Umetsu+2010_CL0024}, so that the
majority of faint blue galaxies are in the far background of typical
cluster lenses,
and that the lensing signal has a weaker
dependence on the source redshift.  In such a case, the single source-plane
approximation may be justified.

\section{Discretized Expressions for the Cluster Lensing Profiles}
\label{appendix:estimators}

\subsection{Averaged Convergence}
\label{appendix:avkappa}

In this appendix, we aim to derive a discrete expression 
for the mean interior convergence $\overline\kappa_\infty(<\theta)$
as a function of clustocentric radius $\theta$
using the azimuthally-averaged convergence $\kappa_\infty(\theta)$.
In the continuous limit,
the mean convergence $\bar\kappa_\infty(<\theta)$ interior to radius $\theta$
can be expressed in terms of $\kappa_\infty(\theta)$ 
as
\begin{equation}
\overline\kappa_\infty(\theta)=\frac{2}{\theta^2}\int_0^{\theta}
\!d\ln\theta'\theta'^2\kappa_\infty(\theta').
\end{equation}
For a given set of $(N+1)$ annular radii $\theta_l$
$(l=1,2,...,N+1)$,
defining $N$ radial bands in the range $\theta_{\rm min}\equiv\theta_1\le
\theta\le \theta_{N+1}\equiv \theta_{\rm max}$,
a discretized estimator for $\overline\kappa_\infty(<\theta)$
can be written in the following way:
\begin{equation}
\label{eq:avkappa_d}
\overline\kappa_\infty(<\theta_l)=
\frac{\theta_{\rm min}^2}{\theta_l^2}\overline{\kappa}_\infty(<\theta_{\rm min})+
\frac{2}{\theta_l^2}\sum_{i=1}^{l-1}
\Delta\ln\theta_i
\bar\theta_i^2
\kappa_\infty(\bar\theta_i),
\end{equation}
with
$\Delta\ln\theta_i \equiv (\theta_{i+1}-\theta_i)/\bar\theta_i$
and $\bar\theta_i$
being the area-weighted center of the $i$th
annulus defined by $\theta_i$ and $\theta_{i+1}$;
in the continuous limit, we have
\begin{equation}
\label{eq:medianr}
\bar\theta_i
\equiv
2\int_{\theta_i}^{\theta_{i+1}}\!d\theta'\theta'^2/
(\theta_{i+1}^2-\theta_{i}^2)\nonumber\\ 
=
\frac{2}{3}
\frac{\theta_{i}^2+\theta_{i+1}^2+\theta_{i}\theta_{i+1}}
{ \theta_{i}+\theta_{i+1} }. 
\end{equation}

\subsection{Lens Distortion and Magnification}

We derive expressions for the binned tangential
distortion $g_{+}(\theta; w_g)$ and magnification $\mu(\theta; w_\mu)$ in terms of the binned
convergence $\kappa_\infty$, using the following relations:
\begin{eqnarray}
g_+(\overline\theta_i; w_g) &=&
\frac{
w_g\left[
\overline{\kappa}_\infty(<\overline\theta_i)-\kappa_\infty(\overline\theta_i)
\right]
}{1-w_g\kappa_\infty(\overline\theta_i)},\\
\mu(\overline\theta_i; w_\mu)&=&\frac{1}{\left[1-w_\mu\kappa_\infty(\overline\theta_i)\right]^2
\left[1-g_+^2(\overline\theta_i; w_\mu)\right]},
\end{eqnarray}
where both the quantities depend on the mean convergence
$\overline\kappa_\infty$ interior to the radius $\overline\theta_i$, which
is the center of the $i$th radial band 
of $[\theta_i,\theta_{i+1}]$ 
(see Appendix \ref{appendix:avkappa}).
By assuming a constant density in each radial band and by noting that
$\overline\theta_i$ is the {\it median} radius of the $i$th
radial band, $\overline\kappa(<\overline\theta_i)$ can be well
approximated by \citep{Umetsu+2011}
\begin{equation} 
\overline{\kappa}_\infty(<\overline\theta_i) = 
\frac{1}{2}\Big[
\overline\kappa_\infty(<\theta_i)+\overline\kappa_\infty(<\theta_{i+1})
\Big],
\end{equation}
where $\overline\kappa_\infty(<\theta_i)$ and $\overline\kappa_\infty(<\theta_{i+1})$
can be computed using the formulae given in this appendix.

Accordingly, all relevant cluster lensing observables,
$g_+(\theta), n_\mu(\theta)$, and $M(<\theta)$,
can be uniquely
specified by the discrete convergence profile,
$\bs=\{\overline\kappa_{\infty,{\rm min}},\kappa_{\infty,i}\}_{i=1}^N$
with $\overline\kappa_{\infty,{\rm min}}\equiv
\overline\kappa_\infty(\theta_{\rm min})$ and $\kappa_{\infty,i}\equiv \kappa_\infty(\theta_i)$.

\end{document}